\documentclass[doc]{apa6}

\usepackage{amsmath}
\usepackage{amsfonts}
\usepackage{amssymb}
\usepackage{listings}
\makeatletter
\def\Let@{\def\\{\notag\math@cr}}
\makeatother
\usepackage{graphicx}
\usepackage[utf8]{inputenc}
\usepackage[T1]{fontenc}

\usepackage[utf8]{inputenc}
\usepackage{amsmath,amssymb}

\usepackage{subcaption}
\usepackage[american]{babel}
\usepackage{csquotes}
\usepackage{natbib}
\usepackage{caption}
\usepackage{{url}}
\usepackage{rotating}

\newcommand\blfootnote[1]{%
  \begingroup
  \renewcommand\thefootnote{}\footnote{#1}%
  \addtocounter{footnote}{-1}%
  \endgroup
}

\title{Estimating Psychopathological Networks: Be Careful What You Wish for}

\shorttitle{CARE IN ESTIMATING PSYCHOPATHOLOGICAL NETWORKS}

\author{Sacha Epskamp, Joost Kruis, Maarten Marsman}

\affiliation{University of Amsterdam: Department of Psychological Methods}

\abstract{Network models, in which psychopathological disorders are conceptualized as a complex interplay of psychological and biological components, have become increasingly popular in the recent psychopathological literature \citep{borsboom2011small}. These network models often contain significant numbers of unknown parameters, yet the sample sizes available in psychological research are limited. As such, general assumptions about the true network are introduced to reduce the number of free parameters. Incorporating these assumptions, however, means that the resulting network will lead to reflect the particular structure assumed by the estimation method---a crucial and often ignored aspect of psychopathological networks. For example, observing a sparse structure and simultaneously assuming a sparse structure does not imply that the true model is, in fact, sparse. To illustrate this point, we discuss recent literature and show the effect of the assumption of sparsity in three simulation studies. }

\begin{document}

\maketitle

\raggedbottom
\urlstyle{same}

\blfootnote{This manuscript is accepted for publication in PlosOne.}

\section*{Introduction}

Recent psychological literature has focused on a network approach to model many different psychological phenomena \citep{schmittmann2013}. Such networks can be high-dimensional structures (i.e., the number of unknown parameters is much larger than the available data), which are hard to estimate without making general assumptions about the underlying true model structure. If the true model is assumed to be sparse, thus containing a small number of connections relative to the number of nodes, a methodology can be applied that potentially returns a sparse network structure. In other words, assuming a sparse network structure results in estimating a sparse network structure, which means that certain conclusions cannot be drawn from observing such a structure. In this paper, we argue that care should be taken in interpreting the obtained network structure because the estimation procedure may pollute the results. We will illustrate this by showing examples of networks obtained when sparse networks are estimated even when the true network structure is dense. 

\section*{Network Psychometrics}

The network approach has been particularly promising in the field of psychopathology. Within this framework, symptoms (e.g., insomnia, fatigue, and concentration problems) are no longer treated as interchangeable indicators of some latent mental disorder (e.g., depression). Instead, symptoms play an active causal role. For example, insomnia leads to fatigue, fatigue leads to concentration problems, and so forth \citep{borsboom2013network}. Psychopathological disorders, then, are not interpreted as the common cause of observed symptoms but rather as emergent behaviors that result from a complex interplay of psychological and biological components. To grasp such a complex structure, a network model can be used in which variables such as symptoms or moods are represented by nodes. Nodes are connected by edges that indicate associations between nodes. This line of research has led to intuitive new insights about various psychopathological concepts such as comorbidity \citep{borsboom2011small,cramer2010comorbidity}, the impact of life events \citep{cramer2012pathoplasticity,fried2015}, and sudden life transitions (e.g., sudden onset of a depressive episode; \citealt{van2014critical, wichers2016critical}).  For an overview of network modeling applied to psychopathology, we refer the reader to a recent review of \citet{fried2017mental}.

	The growing popularity of the network perspective on psychological phenomena has culminated in the emergence of a new branch of psychology dedicated to the estimation of network structures on psychological data---network psychometrics \citep{netpsych}. This field focuses on tackling the problem of estimating network structures involving large numbers of parameters in high-dimensional models. When cross-sectional data are analyzed, the most popular models that are used are the Gaussian Graphical Model (GGM; \citealt{lauritzen1996graphical}) for continuous data and the Ising model \citep{ising1925beitrag} for binary data. Both the GGM and the Ising model fall under a general class of models called \emph{Markov Random Fields}. These models represent variables as nodes which are connected by edges but only if the variables are \emph{conditionally independent}. The strength of an edge (i.e., its absolute deviance from zero) demonstrates the strength of the association between two variables after conditioning on all other variables in the network; this is also termed \emph{concentration} \citep{cox1993linear}. In the GGM, edges directly correspond to partial correlation coefficients. The Ising model does not allow for such standardization, but edge weights can be similarly interpreted. A more detailed introduction of network models is beyond the scope of this paper, but we recommend \citet{netpsych} and \citet{bootnetpaper} for further reading on the subject.

	In these models, we must estimate a weight matrix that contains $P(P-1)/2$ number of parameters, where $P$ is the number of nodes, in order to encode the network structure. These parameters encompass the conditional relationship between two nodes after conditioning on all other nodes in the network and can be shown to be quite instable with relatively low sample sizes \citep{bootnetpaper}. ``Relatively  low sample sizes,'' is a loose description and has not yet been well-defined. A general rule would be to have at least as many observations as the number of parameters. But, as will be shown later, this general rule still results in unstable estimates. A common solution to overcome the problem of estimating many parameters is to reduce this number by using some form of regularization or penalization. A particularly promising technique is to apply the `least absolute shrinkage and selection operator' (LASSO; \citealt{tibshirani1996regression}) to the edge weights of the network. The LASSO penalizes the sum of absolute parameter values such that the estimated values shrink to zero. That is, the absolute parameter estimates will be small and will often equal exactly zero. Therefore, the resulting model is almost always sparse; only a relatively few number of parameters will be estimated to be nonzero. The use of LASSO typically leads to better performance in cross-validations (i.e., overfitting is prevented) and results in more easily interpretable models compared to nonregularized Ising models. Most important is that if the true network structure is sparse, the LASSO performs well in estimating this network structure and, more specifically, in estimating fewer edges to be nonzero that are actually zero in the true network (i.e., fewer false positives). 
	
	The LASSO uses a tuning parameter that controls the sparsity, which can be chosen to minimize some criterion such as the Extended Bayesian Information Criterion (EBIC; \citealt{chen2008EBIC}). This methodology has been shown to work well for both the GGM \citep{foygel2010extended} and the Ising model \citep{foygel2014high,van2014new}, has been implemented in easy-to-use software  \citep{jssv048i04,IsingFit}, and has been utilized in an increasing number of publications \citep{dalege2016toward,isvoranu,kossakowski2015, langley2015should,rhemtulla2016network, van2015association, boschloo2015, levinson2017core}. For a more thorough introduction to this methodology, we recommend reading \citet{primerpaper} and van \citet{van2014new}. 

\section*{Sparse Network Models of Psychopathology}

It has now been routinely observed that network models based on symptoms of different disorders show network structures in which symptoms representative of a disorder strongly cluster together (e.g., \citealt{boschloo2015, bekhuis2016network, beard2016network, levinson2017core}). Commonly, a DSM diagnosis requires an individual to have X out of Y symptoms, regardless of which specific symptoms. This means that two people with vastly different symptoms can be assigned the same diagnosis. This interchangeability results from an underlying causal notion of unobserved diseases causing symptoms rather than symptoms having an active causal role on each other---a notion more formally known as the common cause model \citep{schmittmann2013}. If the common cause model is true, we would expect clustering in the networks much like the clustering found in the literature \citep{netpsych,marsman2015bayesian,kruis2}. These networks, however, do differ on some key aspects as can be expected from interchangeable symptoms: the networks are sparse (contain missing edges), and the number of connections differ per symptom. Furthermore, sometimes negative connections are present where one would expect positive connections. Observing such a structure might lead one to conclude that symptoms are not interchangeable.


	Although we do not necessarily disagree with the notion that symptoms play an active causal role in psychopathology, we wish to point out that the conclusion that symptoms are not interchangeable is difficult to ascertain from a sparse approximated network structure alone. This is because the LASSO relies on the assumption that the true network structure is sparse; the LASSO will always search for a model in which relatively few edges and paths explain the co-occurrence of all nodes. As a result, the LASSO can have a low sensitivity (i.e., not all true edges are detected) but always has a high specificity (i.e., few false positives; \citealt{van2014new}). It is this reason why network analysts prefer the LASSO; edges that are estimated by the LASSO are likely to represent true edges. Moreover, the LASSO returns a possible explanation of the data using only a few connections that can be interpreted as causal pathways \citep{lauritzen1996graphical, pearl2000causality}. That the LASSO yields a possible explanation, however, does not mean that the LASSO provides the only explanation, nor does it indicate that other explanations are false. 
	The sparse explanations found by the LASSO can give great insight regarding a possible way in which psychopathological symptoms interact with each other. However, merely finding a sparse structure does not mean that other explanations (e.g., a common cause with interchangeable symptoms) are disproved. Simply stated, using the LASSO returns a sparse structure, that is what the LASSO does.

\section*{The Bet on Sparsity}

The LASSO is capable of retrieving the true underlying structure but only if that true structure is sparse. Any regularization method makes the assumption that the true structure can be simplified in some way (e.g., is sparse) because otherwise too many observations are needed to estimate the network structure. This principle has been termed the \emph{bet on sparsity} \citep{hastie01statisticallearning}. But what if the truth is not sparse, but dense? 

	Such a case would precisely arise if the true model were a common cause model in which one or several latent variables contribute to scores on completely interchangeable indicators. This is a feasible alternative because the Ising model can be shown to be mathematically equivalent to a certain type of latent variable model: the multidimensional item response model (MIRT; \citealt{reckase2009multidimensional}), with posterior normal distributions on the latent traits \citep{netpsych,marsman2015bayesian}. The corresponding Ising model is a low-rank network that will often be dense (i.e., all possible edges are present). Intuitively, this makes sense because the Ising model parameterizes conditional dependencies between items after conditioning on all other items, and no two items can be made conditionally independent if the common cause model is true. A low-rank weighted network will show indicators of a latent variable as clusters of nodes that are all strongly connected  with each other. Therefore, if a common cause model is the true origin of the co-occurrences in the dataset, the corresponding Ising model should show the indicators to cluster together. Then if LASSO regularization is used, the corresponding network would likely feature sparsity but the nodes would still be clustered together---much like the results in the literature. 
	
	It is this relationship between the Ising model and MIRT that has led researchers to estimate the Ising model using a different form of regularization, by estimating a low-rank approximation of the network structure \citep{marsman2015bayesian}. Such a structure is strikingly different than the sparse structure returned by LASSO estimation. Whereas the LASSO leads to many edge parameters to be exactly zero, a low-rank approximation generally estimates no edge to be exactly zero. Thus a low-rank approximation will typically yield a dense network. On the other hand, this dense network is highly constrained by the eigenvector structure, leading many edge parameters to be roughly equivalent to each other rather than compared to the strongly varying edge parameters LASSO estimation allows. For example, the data can always be recoded such that a Rank 1 approximation only has positive connections. These are key points that cannot be ignored when estimating a network structure. Regardless of the true network structure that underlies the data, the LASSO will always return a sparse network structure. Similarly, a low-rank approximation will always return a dense low-rank network structure. Both methods tackle the bet on sparsity in their own way---sparsity in the number of nonzero parameters or sparsity in the number of nonzero eigenvalues---and both can lose the bet.

\section*{Estimating an Ising Model When the Truth Is Dense}

\begin{figure}[tb]
\centering
\includegraphics[width=1\textwidth]{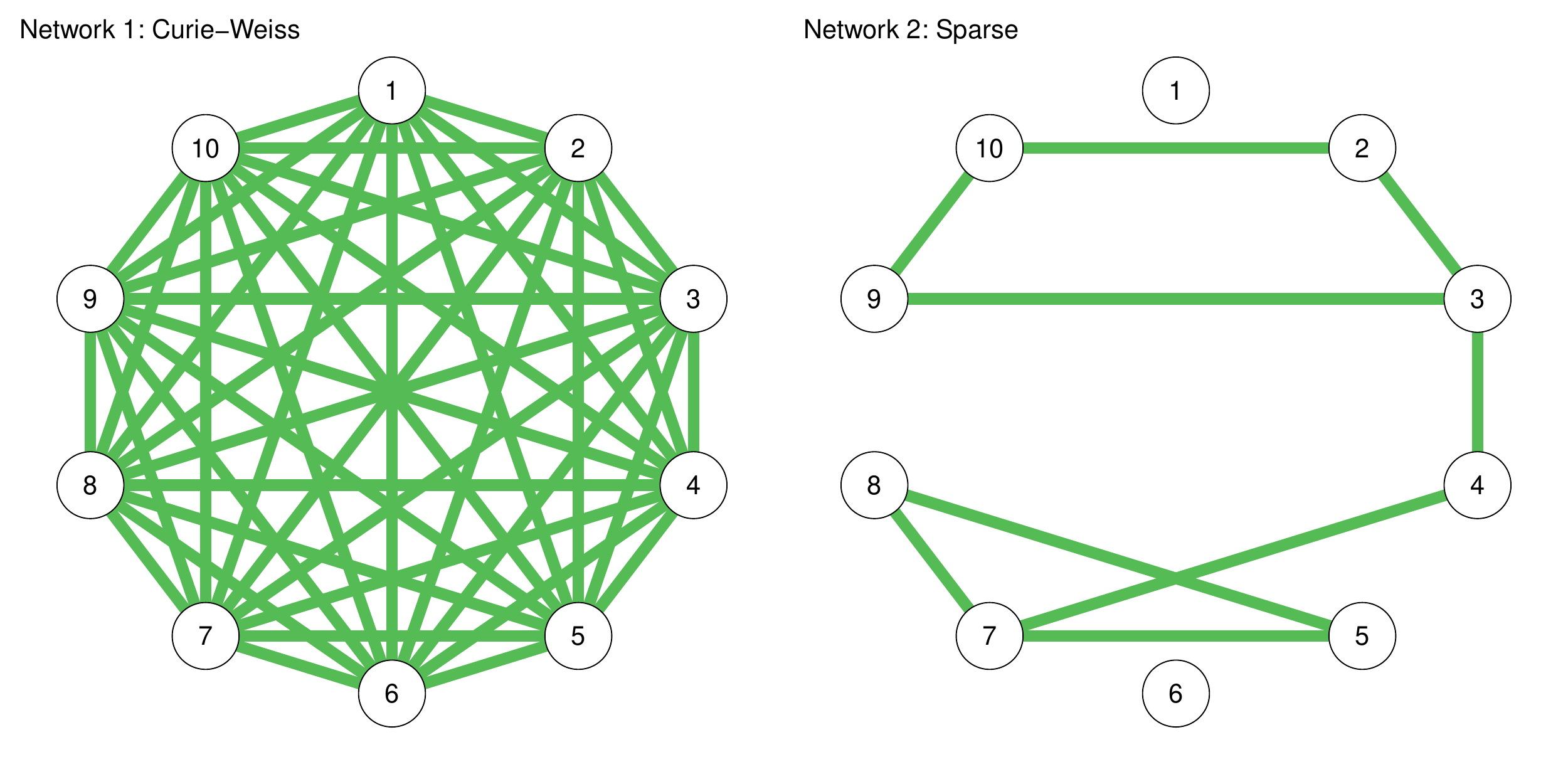}
\caption{True network structures used in simulation study. The first network is a Curie-Weiss network: a fully connected network in which all edges have the same strength. The second network is a random sparse network. All edge weights are $0.2$.}
\label{boschloo:fig:13_1}
\end{figure}

\begin{figure}[tb]
\centering
\includegraphics[width=1\textwidth]{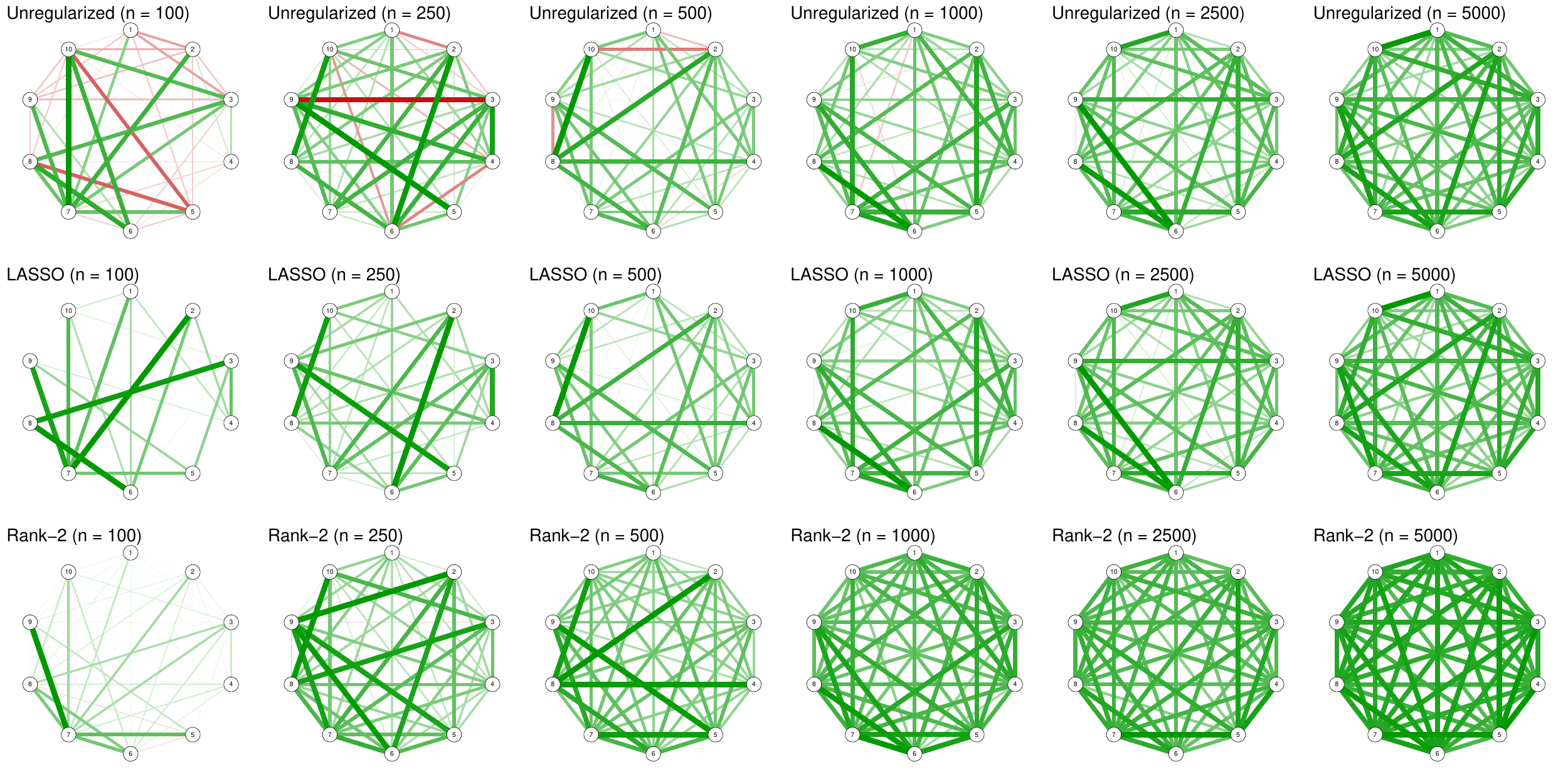}
\caption{Examples of estimated network structures when the true network is a Curie-Weiss network, using different sample sizes and estimation methods. Graphs were drawn using the \emph{qgraph} package without setting a maximum value (i.e., the strongest edge in each network has full saturation and width).}
\label{boschloo:fig:13_2}
\end{figure}

Here we illustrate the effect that the estimation procedure has on the resulting Ising model in two examples. First, we simulated $1{,}000$ observations from the true models shown in Figure~\ref{boschloo:fig:13_1}. The first model is called a Curie-Weiss model \citep{kac1966mathematical}, which is fully connected and in which all edges have the same strength (here set to $0.2$). This network is a true Rank 1 network, which has been shown to be equivalent to a unidimensional Rasch model \citep{marsman2015bayesian}. The Rasch model is a latent variable model in which all indicators are interchangeable. Figure~\ref{boschloo:fig:13_2} shows the results using three different estimation methods---sequential univariate logistic regressions for unregularized estimation \citep{netpsych}, LASSO estimation using the \emph{IsingFit} R package \citep{IsingFit} (all LASSO analyses in this paper make use the default setup of \emph{IsingFit}, using a hyperparameter ($\gamma$) value of $0.25$ as well as the AND-rule), and a Rank~2 approximation \citep{marsman2015bayesian}---on the first $n$ number of rows in the simulated dataset. It can be seen that the unregularized estimation shows many spurious differences in edge strength, including many negative edges. The LASSO performs better but estimates a sparse model in which edge weights vary and in which many edges are estimated to be exactly zero. The Rank~2 approximation works best in capturing the model, which is not surprising because the true model is a Rank~1 network. At high sample sizes, all methods perform well in obtaining the true network structure.

\begin{figure}[tb]
\centering
\includegraphics[width=1\textwidth]{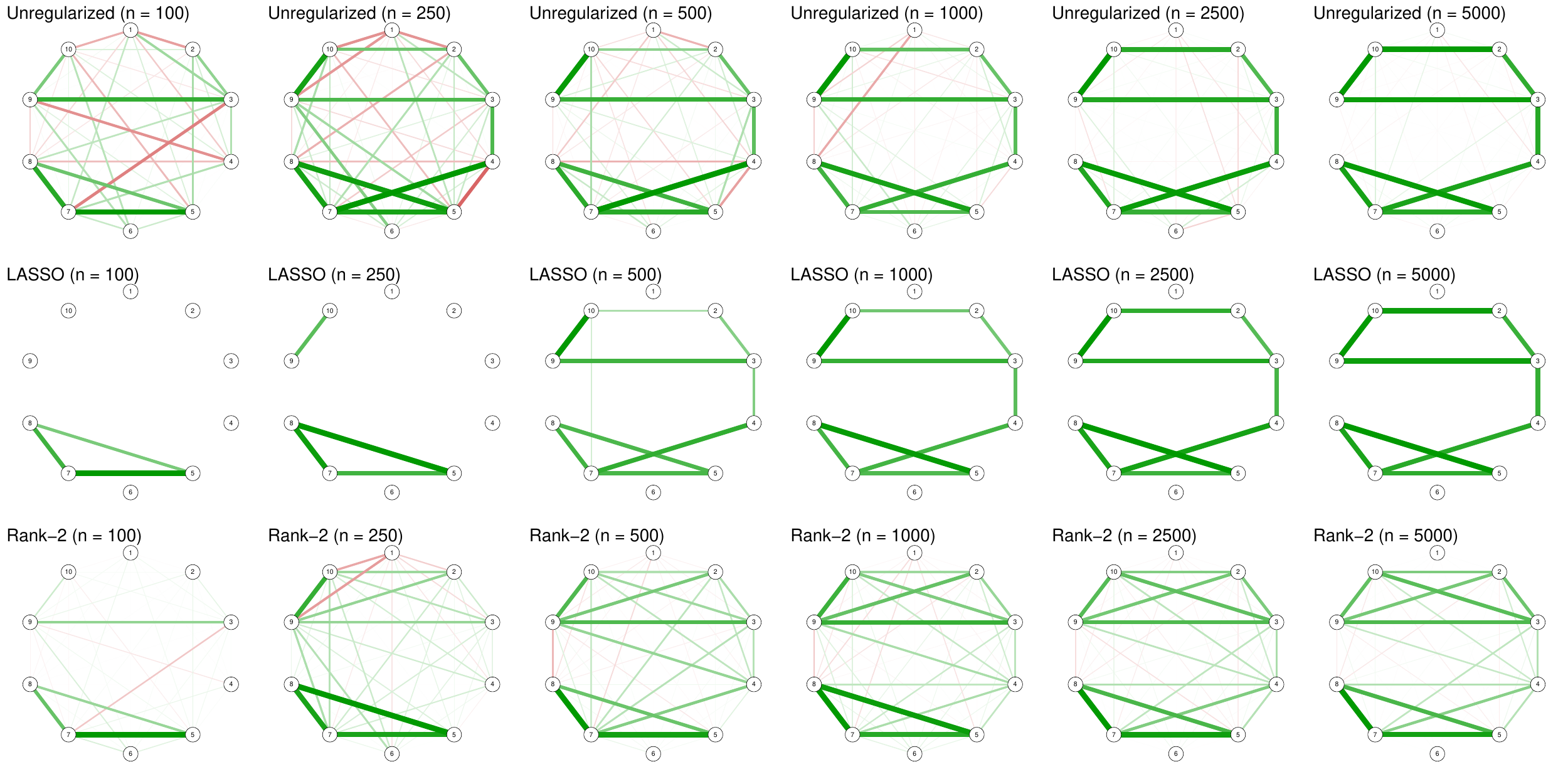}
\caption{Examples of estimated network structures when the true network is sparse, using different sample sizes and estimation methods. Graphs were drawn using the \emph{qgraph} package without setting a maximum value (i.e., the strongest edge in each network has full saturation and width).}
\label{boschloo:fig:13_3}
\end{figure}

The second model in Figure~\ref{boschloo:fig:13_1} corresponds to a sparse network in which $20\%$ of randomly chosen edge strengths are set to $0.2$ and in which the remaining edge strengths are set to 0 (indicating no edge). As Figure~\ref{boschloo:fig:13_3} shows, the LASSO now performs very well in capturing the true underlying structure. Because both the unregularized estimation and the Rank~2 approximation estimate a dense network, they have a very poor specificity (i.e., many false-positive edges). In addition, the Rank~2 approximation retains spurious connections even at high sample sizes (choosing a higher rank will lead to a better estimation). Thus, this example serves to show that the LASSO and low-rank approximations only work well when the assumptions on the true underlying model are met. In particular, using a low-rank approximation when the truth is sparse will result in many false positives, whereas using a LASSO when the truth is dense will result in many false negatives. Even when the true model is one in which every node represents an interchangeable symptom, the LASSO would still return a model in which nodes could be interpreted to not be interchangeable.

\begin{figure}[tb]
\centering
\includegraphics[width=1\textwidth]{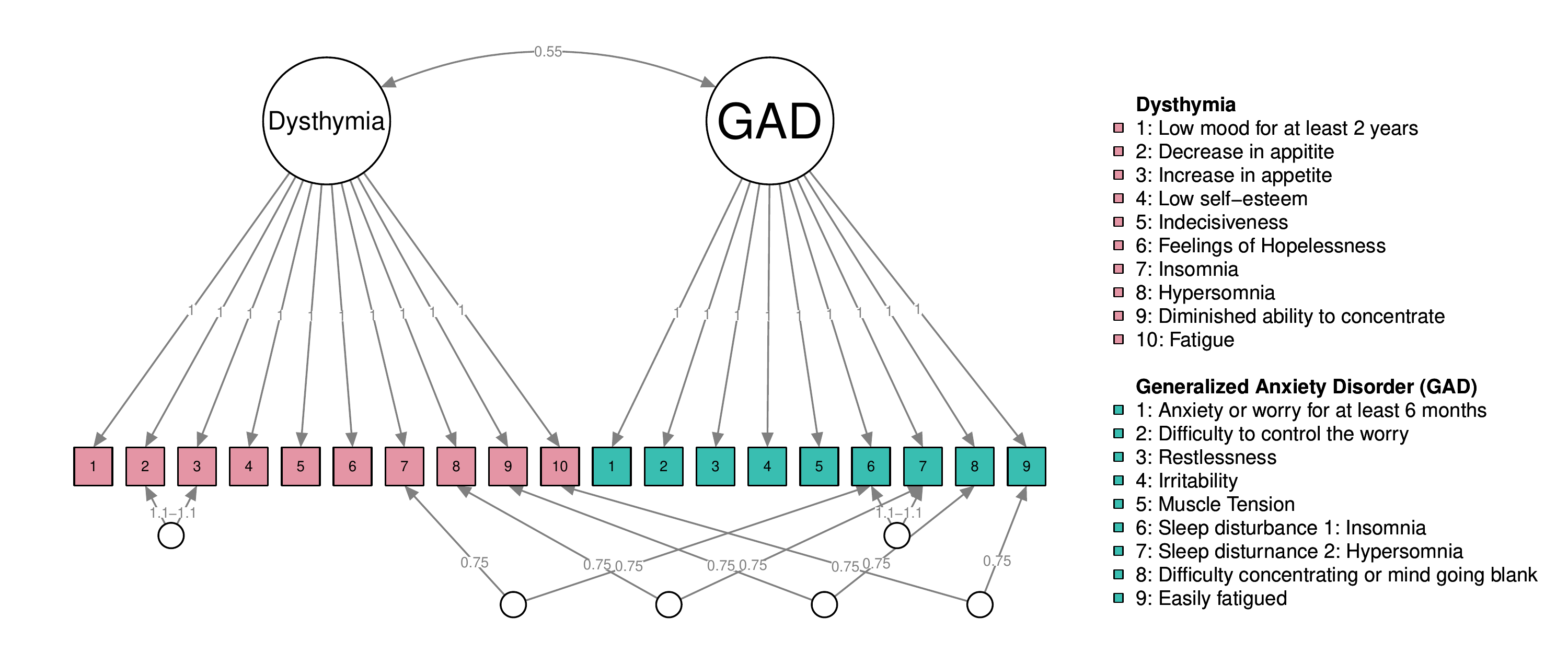}
\caption{A multidimensional IRT model (MIRT) used in simulating data. All latent variables were normally distributed with standard deviation of 1 and all symptoms were binary. The edges in this model correspond to item discrimination parameters.}
\label{boschloo:fig:13_4}
\end{figure}

	For the second example, we simulated data under the latent variable model as shown in Figure~\ref{boschloo:fig:13_4}, using an MIRT model \citep{reckase2009multidimensional}. In this model, the symptoms for dysthymia and generalized anxiety disorder (GAD) were taken from the supplementary materials of \citet{boschloo2015}, with the exception of the GAD symptom ``sleep disturbance,'' which we split in two: insomnia and hypersomnia. The item discriminations of each symptom were set to 1, indicating that symptoms are interchangeable, and item difficulties were set to 0. All latent variables were simulated to be normally distributed with a standard deviation of 1, and the correlation between dysthymia and GAD was set to 0.55---similar to the empirically estimated comorbidity \citep{kessler2005prevalence}. Nodes 2 and 3 in dysthymia and nodes 6 and 7 in GAD are mutually exclusive, which we modeled by adding orthogonal factors with slightly higher item discriminations of 1.1 and -1.1. Furthermore, nodes 7, 8, 9, and 10 of dysthymia are identical to nodes 6, 7, 8, and 9 of GAD respectively, which we modeled by adding orthogonal factors with item discriminations of 0.75. These nodes may not be identical because a skip structure can be imposed on the questionnaire (e.g., \citealt{boschloo2015, borsboom2013network}). That is, if someone does not exhibit the symptom ``low mood,'' that person is never asked about insomnia in the depression scale because he or she is assumed to not have this symptom. We did not impose a skip structure to keep the simulation study simple. Such shared symptoms are termed \emph{bridge symptoms} in network analysis because they are assumed to connect the clusters of disorders and explain comorbidity \citep{borsboom2011small,cramer2010comorbidity}. In sum, the model shown in Figure~\ref{boschloo:fig:13_4} generates data that are plausible given the latent disease conceptualization of psychopathology.
	
\begin{figure}
\centering
\includegraphics[width=1\textwidth]{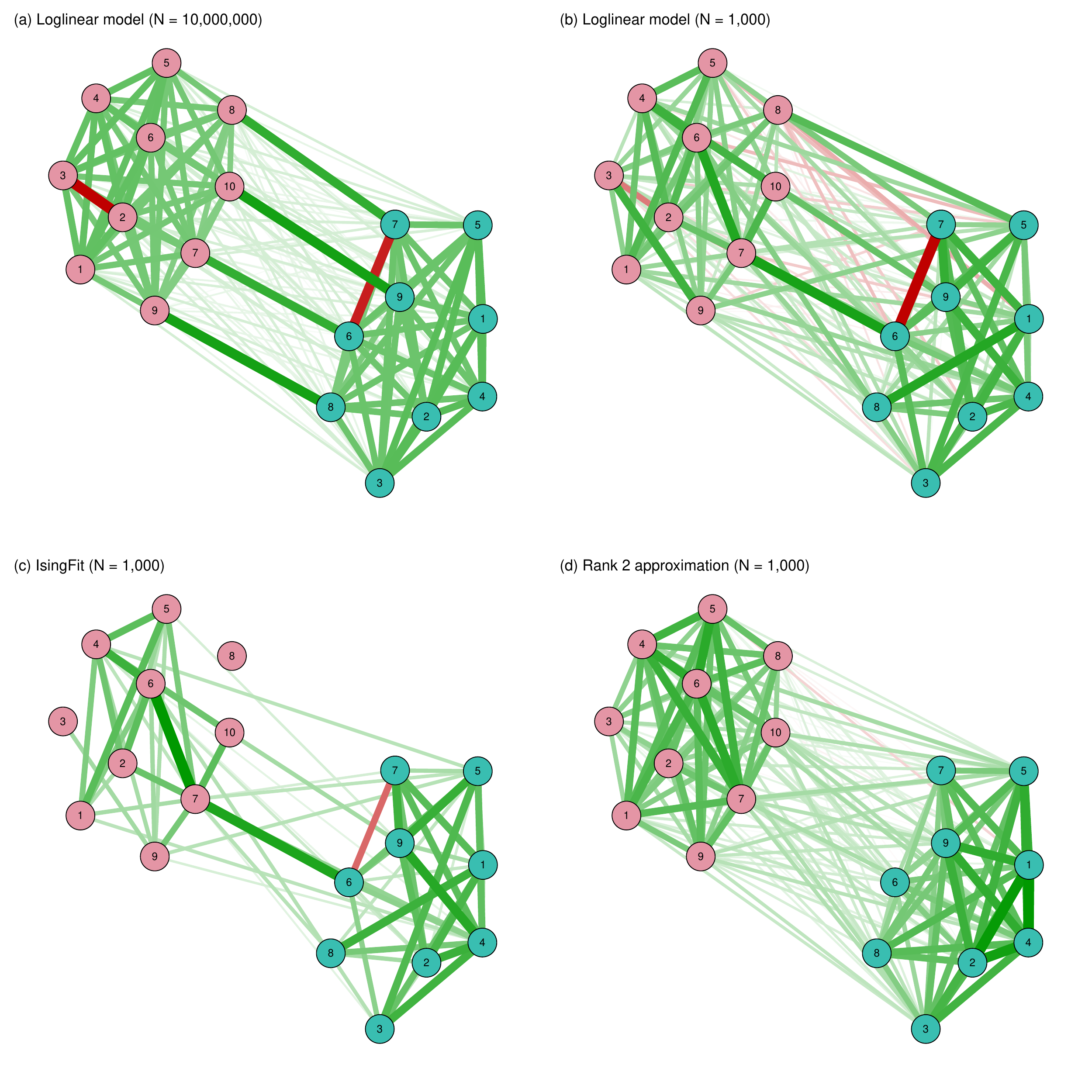}
\caption{Estimated network structures based on data generated by the MIRT model in Figure~\ref{boschloo:fig:13_4}.}
\label{boschloo:fig:13_5}
\end{figure}

	Figure~\ref{boschloo:fig:13_5} shows the simulated and recovered network structures. First we simulated $10$ million observations from this model and estimated the corresponding Ising model using nonregularized estimation by framing the Ising model as a log-linear model \citep{agresti2014categorical,netpsych} (the estimation was done using the \emph{IsingSampler} package; \citealt{IsingSampler}). Panel~A shows the results, which give a good proxy of the true corresponding Ising structure. It can be seen that the true model is dense, meaning that indicators of the disorders cluster together. Two negative connections are formed between the mutually exclusive indicators, and bridging connections are formed between the shared indicators. Next, we simulated $1{,}000$ observations from the model in Figure~\ref{boschloo:fig:13_4} and estimated the Ising model in various ways. Panel~B shows unregularized estimation via a log-linear model and shows many spurious strong connections, including many more negative connections than present in the true model. As such, Panel~B highlights our need to regularize---even in a sizable dataset of $1{,}000$ observations for a 19-node network. The simulated data has $22.2$ observations for every parameter,
	Thus, even with a high sample size and even when more subjects are measured than there are parameters present, it can still be advisable to use some form of regularization.
	Panel~C shows the result from using the LASSO, using the \emph{IsingFit} package \citep{van2014new}. In this model, the clustering is generally retrieved---two of the bridging connections are retrieved and one negative connection is retrieved. However, the resulting structure is much more sparse than the true model, and interpreting this structure could lead one to conclude that the number of connections differed across symptoms, connection strengths varied considerably across symptoms, and relatively few connections connected the two disorders. Finally, Panel~D shows the result of a Rank 2 approximation, which is equivalent to a two-factor model. Here, it can be seen that although a dense structure is retrieved that shows the correct clustering, violations of the clustering (the negative and bridging edges) are not retrieved. The supplementary materials show that with a higher sample size ($n = 5{,}000$) the estimation is more accurate and that the unregularized and LASSO estimations result in similar network structures.

\subsection*{Different Estimation Techniques}

In light of the examples discussed in this paper, researchers may wonder when they should and should not use a particular estimation method. For example, low-rank estimation is more suited in the example demonstrated in Figure~\ref{boschloo:fig:13_2}, whereas LASSO estimation fits better in the example shown in Figure~\ref{boschloo:fig:13_3}. These conclusions, however, depend on knowing the true network structure as shown in Figure~\ref{boschloo:fig:13_1}---something a researcher will not know in reality. The choice of estimation method, therefore, is not trivial. Choosing the estimation method depends on three criteria: (1) the prior expectation of the true network structure, (2) the relative importance the researcher attributes to sensitivity (discovery) and specificity (caution), and (3) the practical applicability of an estimation procedure. When a researcher expects the true network to be low rank (e.g., due to latent variables), low-rank estimation should be preferred over LASSO regularization. On the other hand, when a researcher expects the network to be sparse, LASSO regularization should be used. In addition, LASSO regularization should be preferred when a researcher aims to have high specificity (i.e., to refrain from estimating an edge that is missing in the true model). Finally, practical arguments can play a role in choosing an estimation procedure as well. LASSO, particularly in combination with EBIC model selection, is relatively fast even with respect to large datasets. As a result, researchers could apply bootstrapping methods to the estimation procedure to further investigate the accuracy of parameter estimation \citep{bootnetpaper}, which may not be feasible for slower estimation procedures.

We focused the argumentation on LASSO regularization and low-rank approximation because these are the main methodologies that have been applied in psychological literature and present two extreme cases of a range of different network structures that can be estimated. Because these methods lie on the extreme ends of sparsity relative to dense networks, they best exemplify the main point of this paper: In small sample sizes, some assumptions of the true model must be made (e.g., the true model is sparse), and these assumptions influence the resulting network structure (e.g., the obtained network is sparse). This does not mean that LASSO and low-rank approximation are the only methods available.  An alternative, for example, is to use \emph{elastic-net} estimation, which mixes LASSO regularization with ridge regression (penalizing the sum of squared coefficients). The \emph{elasticIsing} package \citep{elasticising} can be used to accomplish this; it uses cross-validation in selecting the tuning parameters. The supplementary materials show an example of elastic-net applied to the data analyzed in Figure~\ref{boschloo:fig:13_5}. It is noteworthy that the elastic-net procedure selected a dense network (i.e., ridge regression) over LASSO regularization, indicating that data-driven evidence can be garnered to argue whether or not LASSO regularization should be used. The obtained network, like the unregularized network in Figure~\ref{boschloo:fig:13_5} (Panel~B), also shows many connections which were falsely estimated to be negative; this raises the question of whether its result should or should not be preferred over LASSO regularized estimation. The supplementary materials also contain examples of LASSO regularization using different tuning arguments (e.g., BIC selection instead of EBIC selection), which improves sensitivity (i.e., more edges are detected) in this particular case. Doing so, however, will result in less specificity when the true model is sparse \citep{van2014new}. Finally, promising methodology has been proposed to combine latent variable and network modeling, allowing one to combine sparse and low-rank network approximation \citep{epskampPsychometrika,fusedlasso,chandrasekaran2010latent,panPsychMethods}.

\section*{Conclusion}

Network estimation has grown increasingly popular in psychopathological research. The estimation of network structures, such as the Ising model, is a complicated problem due to the fast growing number of parameters to be estimated. As a result, the sample size typically used in psychological science may be insufficient to capture the true underlying model. Although a large sample size network estimation typically goes well regardless of the estimation method used (see supplementary materials), Figures \ref{boschloo:fig:13_2}, \ref{boschloo:fig:13_3}, and \ref{boschloo:fig:13_5} show that estimating an Ising model with sample sizes commonly used in psychological research results in poor estimates without the use of some form of constraint on the parameter space. Two such constraints involve limiting the size and number of nonzero parameters (LASSO) or reducing the rank of a network (low-rank approximation). It is important to realize that using such estimation methods makes an assumption on the underlying true model structure: The LASSO assumes a sparse structure whereas low-rank approximation assumes a dense but low-rank structure. Investigating the results of the estimation methods cannot validate these assumptions. The LASSO always yields a sparse structure, which does not mean that the true underlying structure could not have been dense. On the other hand, low-rank approximations rarely produce sparse structures, but that does not mean that the true underlying structure could not have been sparse.

	Figure~\ref{boschloo:fig:13_2} illustrates this point by showing that LASSO estimation when the true network structure is a Curie-Weiss model still results in a sparse structure. This means that observing any of the sparse structures shown in Figure~\ref{boschloo:fig:13_2} does not mean that the nodes in the network could not represent interchangeable indicators of a single latent trait. Figure~\ref{boschloo:fig:13_5} illustrates this point again in a plausible scenario in psychopathology and also shows that when the true network structure is complicated and neither sparse nor low rank, as is the case here, all regularization methods partly fail even when using a relatively large sample size. As such, interpreting the sparsity of such a structure is questionable; the LASSO resulting in a sparse model gives us little evidence for the true model being sparse because a low-rank approximation returning a dense model seems to indicate that the true model is dense. Those characteristics from the networks we obtain are a consequence of the method used to estimate a network structure (specifically the assumptions made by the employed method about the data-generating network structure) and often pollute the resulting estimated model \citep{kruisThesis}. 
	
Recently it has been demonstrated that three, statistically indistinguishable, representations of the Ising model exist that explain observed associations between binary variables either through a common cause (latent variable), through the reciprocal effect between variables (network), or through the conditioning on a common effect (collider variable) \citep{netpsych,marsman2015bayesian,kruis2}. Consequently, when a model from one of these frameworks can sufficiently describe the associative structure of the measured variables, there exists an alternative representation for other frameworks that can also accurately represent the structure of the data. 
For example, spare network structures \citep{boschloo2015, bekhuis2016network, beard2016network, levinson2017core}, resulting from the LASSO being applied to the data, can also be described by a multidimensional latent variable model (with a single latent variable for each clique in the network) and residual correlations. As such, obtaining sufficient fit for a statistical network model cannot be regarded as evidence for the theoretical model, where a network structure acts as the causal mechanism from which associations between variables emerge. We therefore advise, in general, to tread carefully when drawing inferences about the theoretical causal mechanisms that generate the data from statistical model fit.

Network models show great promise in mapping out and visualizing relationships present in the data and are useful to comprehend high-dimensional multivariate relationships. In addition, network models can be powerful tools to estimate the backbones of potential causal relationships---if those relationships are assumed to exist. Using the LASSO to estimate such network structures is a powerful tool in performing fast high-dimensional model selection that results in fewer false positives, and interpreting network structures obtained from the LASSO can illuminate the strong relationships present in the dataset. Important to realize is that using LASSO estimation will result in a sparse structure, and similarly, using a low-rank approximation will result in a dense low-rank result. Our aim here is not to argue against using the LASSO or to argue that estimating network structures is wrong. Our aim is to clarify that choosing the estimation method is not trivial and can greatly impact both the estimated structure as well as any conclusions drawn from that structure.

\section*{Acknowledgements}

We would like to thank Lynn Boschloo for constructive feedback on earlier versions of our manuscript.

\bibliographystyle{apalike}
\bibliography{Bibliography}

\end{document}